\documentclass[12pt]{article}
\pdfoutput=1

\usepackage{epsfig}
\usepackage{amsfonts}
\usepackage{amssymb}
\usepackage{xcolor}

\begin{document}
\newcommand{\nc}{\newcommand}
\nc{\beq}{\begin{equation}} \nc{\eeq}{\end{equation}}
\nc{\beqa}{\begin{eqnarray}} \nc{\eeqa}{\end{eqnarray}}
\nc{\R}{{\cal R}}
\nc{\A}{{\cal A}}
\nc{\K}{{\cal K}}
\nc{\B}{{\cal B}}
\begin{center}

{\bf \Large  NON-RENORMALIZABLE INTERACTIONS:\\[0.4CM]  A SELF-CONSISTENCY MANIFESTO} \vspace{1.0cm}

{\bf \large D. I. Kazakov} \vspace{0.5cm}

{\it
Bogoliubov Laboratory of Theoretical Physics, Joint
Institute for Nuclear Research, Dubna, Moscow region, Russia.}
\vspace{0.5cm}

\abstract{The renormalization procedure is proved to be a rigorous way to get finite answers in  a renormalizable class of field theories. We claim, however, that it is redundant if one reduces the requirement of finiteness to S-matrix elements only and does not require finiteness of intermediate quantities like the off-shell Green functions. We suggest a novel view on the renormalization procedure. It is based on the usual BPHZ $\R$-operation, which is equally applicable  to any local QFT independently of whether it is renormalizable or not.  The key point is the replacement of the multiplicative renormalization, used in renormalizable theories, by an operation when the renormalization constants depend on the fields and momenta that have to be integrated inside the subgraphs.
This approach being applied to quantum field theories does not distinguish between  renormalizable and non-renormalizable interactions and  provides the basis for getting finite scattering amplitudes in both cases. 
The arbitrariness of the subtraction procedure is fixed by imposing a normalization condition on the scattering amplitude as a whole rather than on an infinite series of new operators appearing in the process of subtraction of UV divergences in non-renormalizable theories.

We show that  using the property of locality of counter-terms, precisely as in renormalizable theories, one can get recurrence relations connecting leading, subleading, etc., UV divergences in all orders of perturbation theory in any local theory. This allows one to get
 generalized RG equations that have an integro-differential form and sum up the leading  logarithms in all orders of PT  in full analogy  with the renormalizable case. This way one can cure the problem of violation of unitarity in non-renormalizable theories by summing up the leading asymptotics.  
  The approach can be applied to any theory though technically non-renormali\-zable interactions are much more complicated than renormalizable ones. We illustrate the basic features  of our approach by several examples.
 
 Our main statement is that non-renormalizable theories are self-consistent, they can be well treated within the usual BPHZ  $\R$-operation, and the arbitrariness can be fixed  to a finite number of parameters just as in the renormalizable case. }

\end{center} 

\section{Introduction}

The classification of local quantum field theories into renormalizable and non-renormali\-zable ones  is based on the analysis of ultraviolet (UV) divergences. In renormalizable theories, local counter-terms that eliminate UV divergences repeat the structure of the original Lagrangian and, therefore, can be absorbed into renormalization of the cor\-responding terms in the Lagrangian. In non-renormalizable theories, on the contrary, in each order of perturbation theory, new structures appear that do not repeat the original ones, and, therefore,  they cannot be absorbed into renormalization.  Thus, in the first case, the Lagrangian, including the counter-terms, contains a finite number of structures and is closed in terms of renormalizations, and, in the second case, there is an infinite number of terms, although limited in each given order of perturbation theory. The latter situation is usually considered as unacceptable, because the standard procedure prescribes normali\-za\-tion of each newly appearing term, the number of which is infinite, and, accordingly, there is  infinite arbitrariness in the choice of parameters.

We propose here to look at this situation somewhat differently and do not require to fix  arbitrariness in each new term  but to fix the entire expression for the amplitude as a whole.
This reduces the arbitrariness of the subtraction procedure to a finite set of couplings just as in the renormalizable case. The difference is that the coupling constants do not belong to a single operator but 
to the whole infinite sequence of operators with increasing number of derivatives and fields.
At the same time, the usual BPHZ $\R$-operation\cite{BP,BPHZ} aimed at eliminating UV divergences remains unchanged. Divergences are still eliminated by the introduction of local counter-terms, however, unlike the renormalized case, this procedure is not equivalent to the multiplicative renormalization of  the Green functions and parameters, but corresponds to a more complex operation when the renorma\-lization constant ceases to be a constant but depends on kinematic variables and fields. 
Thus, renormalizing the coupling, we simultaneously renormalize an infinite set of operators with the same coupling. The renormalizable class of theories in this approach happens to be the simplified case, but the procedure equally works in all theories.
Below, we describe this procedure and show how it works order by order of PT. Based on the Bogolyubov-Parasyuk theorem~\cite{BP} on the locality of counter-terms, we also obtain recurrence relations that link the divergences in subsequent  orders of PT, which allows us to find leading divergences in any order of PT based on the one-loop expressions. These recurrence relations are then converted into differential equations for the total sum of  PT series, which serve as a generalization of the renormalization  group equations to the non-renormalizable case. We consider solutions of these equations in some models below. They allow us to perform the summation of the leading asymptotics and this way to check the unitarity and  the UV-completion of a given theory.

\section{Momentum and field dependent renormalization}

Following the procedure of $\R $ -operation of Bogolyubov-Parasyuk-Hepp-Zimmermann\cite{BPHZ,Rop}, the elimination of UV divergences in any local theory is achieved by  introducing local counter-terms. To ensure locality, it is necessary to introduce the counter-terms sequentially order by order of perturbation theory and take into account the previously introduced ones to eliminate divergences in subgraphs. Since the counter-terms are local, the introduction of a lower-order counter-term into the diagram corresponds to shrinking the corresponding divergent subgraph to a point and multiplying the diagram obtained in this way by a factor equal to the coefficient of this counter-term. In a renormalizable theory, this coefficient is a constant and, therefore, the described operation is simply a multiplication, which further leads to a multiplication of the corresponding amplitude by a constant factor $Z$ called the renormalization constant. 

In non-renormalizable theories, the coupling constant has a negative mass dimension, so to compensate for the dimension, the counter-terms are proportional to the powers of  momenta (and/or fields), which corresponds to higher derivatives in the coordinate representation or extra field operators. The dependence of the counter-terms on momentum means that in the above procedure  after shrinking the subgraph to a point, the momenta must be integrated inside the remaining reduced diagram. Thus, the described procedure is not a simple multiplication operation anymore~\cite{K1}. 

Let us show by simple examples how this operation works.  Note that though our main statements are valid in any theory and in any regularization, each particular case has its specific features. Below, we  apply dimensional regularization and  calculate the diagrams in dimension $D-2\epsilon$, where $D$ is an integer and $ \epsilon \to 0$. 

\subsection{$g \phi^4_D$ Theory} 
\begin{figure}[h]
\begin{center}
\includegraphics[scale=0.60]{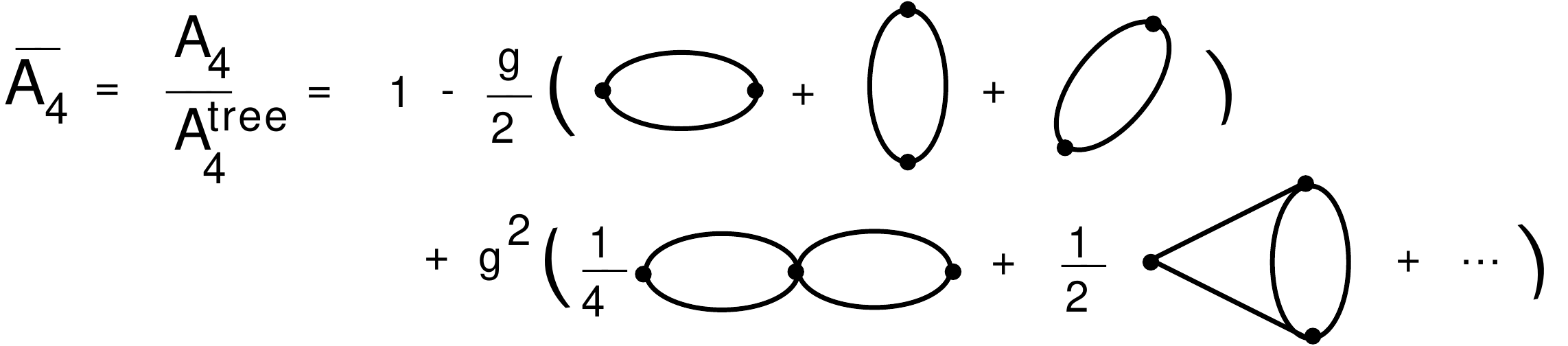}
\caption{The one- and two-loop diagrams contributing to the four-point amplitude \label{2loopd}}
\end{center} 
\end{figure}

Take the scalar field theory with the interaction $g \phi^4$ in $D$ dimensions and consider the four-point amplitude on mass shell~\cite{K2}. The corresponding one- and two-loop diagrams are shown in Fig.\ref{2loopd}.

\noindent The divergence of the  one-loop diagram is \ $ \sim \frac{1}{\epsilon}(p^2)^{D/2-2},$ \ where $p$ is the momentum flowing into the diagram. Hence, in four dimensions there is no momentum dependence, and in higher dimensions, we have a polynomial dependence on $p^2$ (we consider even values of $D$). To eliminate this divergence, a counter-term of the following form is introduced into the Lagrangian
\beq
\Delta {\cal L} \sim-g^2\frac{1}{\epsilon}(p^2)^{D/2-2}\phi^4,
\eeq
that in the coordinate space corresponds to the expression
\beq
\Delta {\cal L} \sim-g^2\frac{1} {\epsilon}(\partial^{D/2-2}\phi^2)^2.
\eeq

\begin{figure}[h]
\begin{center}
\includegraphics[scale=0.50]{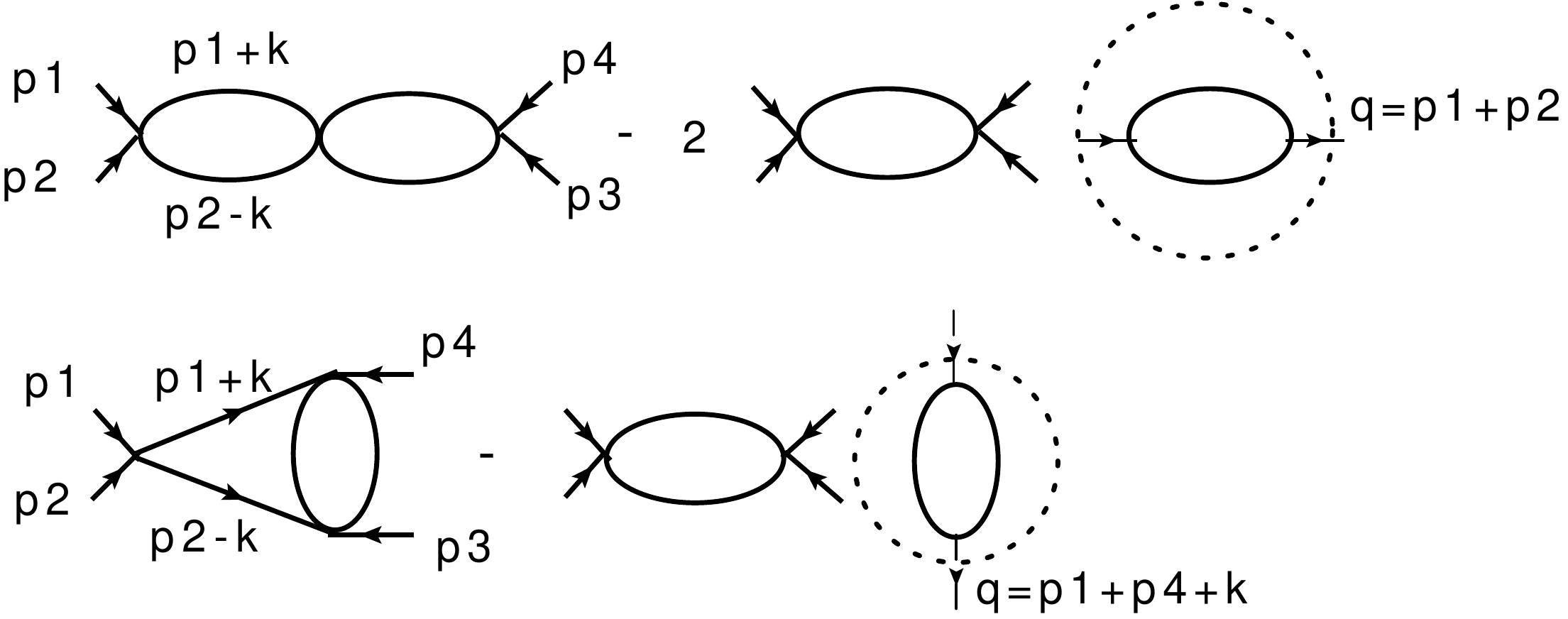}
\caption{Subtraction of subdivergences in  the two-loop diagrams \label{R2loops}}
\end{center}
\end{figure}

When removing divergences from the  two-loop diagram, this counter-term must be taken into account, which leads to the subtraction of the divergences in the subgraph (see Fig.\ref{R2loops}).

For $D=4$, the coefficient for the counter-term is a constant, and in the last terms we get a simple multiplication of the one-loop subgraph obtained after shrinking the divergent subgraph to a point, by this constant. However, for $D>4$, the counter-term depends on momentum $q$ flowing through the divergent subgraph. In the first line, this momentum $q=p_1+p_2$  is fixed  and one has the usual multiplication, while  in the second line $q=p_1+p_4+k$ and has to be integrated over internal momentum $k$ in the remaining subgraph. 
Thus, in this case, the operation of eliminating the divergence in the subgraph is not reduced to a simple multiplication.

However, formally, the ${\cal R}$ - operation, even in the non-renormalizable case, can be formulated as multiplying the amplitude $ A_4$ by the renormalization constant $Z_4$, which depends on kinematic variables and fields and acts as an operator, and the corresponding renormali\-za\-tion of the coupling constant $g$~\cite{K1}  
\beqa
\A_4& = &Z_4(g) \A_4^{bare}|_{g_{bare}->gZ_4}, \label{mult}\\
g_{bare}&=&\mu^ \epsilon Z_4(g)g. \label{coupling}
\eeqa
The renormalization constant 
 $Z_4$ is calculated diagrammatically using the standard formula~\cite{Vas}:
\beq
Z = 1-\sum_i\K\R' G_i, \label{ZZ}
\eeq
where the sum goes over all divergent subgraphs.
The incomplete ${\cal R}$ - operation (${\cal R}^\prime$-operation) subtracts only the subdivergences in the graph, and the full ${\cal R}$ - operation is defined by the relation
\begin{equation}
\R G = (1-\K) \R' G.
\end{equation}
Here the operator ${\cal K}$ retrieves the singular part of the diagram, and ${\cal KR}^\prime G$ is a counter-term corresponding to the graph $G$~\cite{Vas}.  

Let us see how it works in the above example. 
The one loop renormalization constant in this case is 
\beq
Z_4^{(1)}=1+g \frac{1}{2\epsilon}(s^{D/2-2}+t^{D/2-2}+u^{D/2-2}),
\eeq
where $s,t$ and $u$ are the usual Mandelstam variables.
Multiplying the amplitude $A_4$ by $Z_4$ and replacing the bare coupling $g$ according to eq.(\ref{coupling}),  one has in the order of $g^2$ (in the s-channel)\\

\includegraphics[scale=0.50]{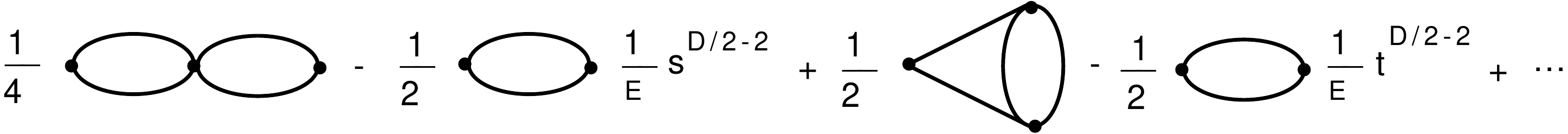}\\

This way we reproduce the $\R$-operation for the two-loop diagrams. Note that while multiplication by $s^{D/2-2}$ is a simple multiplication, multiplication by  $t^{D/2-2}$ should be understood as  integration over the one-loop subgraph.

Thus, one comes to the notion of {\it kinematically dependent renormalization} \cite{K1} when the renormalization constant $Z$ is not a constant anymore but depends on kinematics. These kinematic factors have to be integrated through the lower loop diagrams. 
The multiplication of  the coupling constant by a factor $Z$ depending on the kinematics should be understood as generating a new vertex with higher derivatives. 

\begin{figure}[h]
\begin{center}
\includegraphics[scale=0.50]{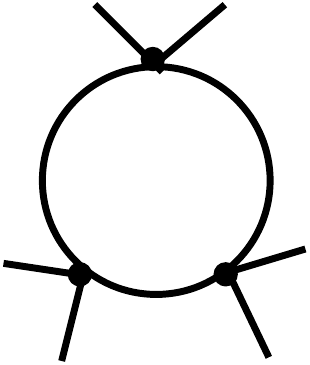}
\caption{The one loop six-leg diagram \label{6leg}}
\end{center}\end{figure}

Consider now the amplitude with 6 legs. The simplest one-loop diagram is shown in Fig.\ref{6leg}.
This diagram is convergent for $D=4$ but diverges for $D>4$. The corresponding counter-term
is
\beq
\Delta {\cal L} \sim-g^3\frac{1}{\epsilon}(p^2)^{D/2-3}\phi^6,
\eeq
or in the coordinate space
\beq
\Delta {\cal L} \sim-g^3\frac{1} {\epsilon}(\partial^{D/2-3}\phi^2)\phi^4.
\eeq
This counter-term  is also absent in the original Lagrangian and hence cannot be absorbed into the renormalization constant of the $\phi^4$ operator. However, following the previous procedure with the higher derivative terms, one can include this counter-term into the {\it field-dependent} renormalization constant. This way we collect not only terms with higher derivative operators but also terms with an extra number of fields. All these terms contribute to the four-point amplitude at a certain loop order.

Continuing this procedure in higher orders, we get the following infinite sequence of terms contributing to $Z_4$:
\beqa
Z_4&=&1 +\frac{g} {\epsilon}(p^2)^{D/2-2}+\frac{g^2} {\epsilon^2}(p^4)^{D/2-2}+\frac{g^3} {\epsilon^3}(p^6)^{D/2-2}+\dots  \ \ \leftarrow \mbox{leadimg order} \nonumber\\
&&\hspace{3cm}+\frac{g^2} {\epsilon}(p^4)^{D/2-2}+\frac{g^3} {\epsilon^2}(p^6)^{D/2-2}+\dots  \ \ \leftarrow \mbox{subleadimg order} \nonumber\\
&&  \hspace{5.6cm}+ \ \dots \nonumber \\
&& \hspace{3cm}+\frac{g^2} {\epsilon}(p^2)^{D/2-3}\phi^2+\dots \hspace{2.5cm} \ \ \ (D\geq 6)\nonumber\\
&& \hspace{5.5cm}+\frac{g^2} {\epsilon}(p^2)^{D/2-4}\phi^4+\dots \ \ \ \ (D\geq 8)\nonumber\\
&&  \hspace{8.2cm}+ \ \dots  \label{Z}
\eeqa

Equation (\ref{Z}) is written in a symbolic form since the momentum dependence in multi-leg diagrams might be more complicated and correspond to various ways to put the derivatives in the coordinate space.
This way we generate an infinite sequence of vertices with higher derivatives and larger number of legs.
Thus, the coupling constant $g$ becomes not just the coefficient of a single operator but also of an infinite series of terms, and when renormalizing the coupling, we do not renormalize a single one but the whole series.

\subsection{SYM$_D$ Theory}

As another example,  we consider the maximally supersymmetric Yang-Mills theory.
  To calculate the amplitude, it is convenient first to extract the colour ordered partial amplitude by executing the colour decomposition~\cite{Reviews_Ampl_General}
\begin{equation}
\mathcal{A}_n^{a_1\dots a_n,phys.}(p_1^{\lambda_1}\dots p_n^{\lambda_n})=\sum_{\sigma \in S_n/Z_n}Tr[\sigma(T^{a_1}\dots T^{a_n})]
\mathcal{A}_n(\sigma(p_1^{\lambda_1}\dots p_n^{\lambda_n}))+\mathcal{O}(1/N_c).
\end{equation}

The colour ordered amplitude $A_n$ is evaluated in the limit $N_c\to \infty$, $g^2_{YM}\to 0$ and $g^2_{YM}N_c$ is fixed, which corresponds to the planar diagrams.
In the case of the four-point amplitudes, the colour decomposition is performed as follows:
\begin{eqnarray}
\mathcal{A}_4^{a_1\dots a_4,(L),phys.}(1,2,3,4)=T^1\mathcal{A}_4^{(L)}(1,2,3,4)+T^2\mathcal{A}_4^{(L)}(1,2,4,3)+
T^3\mathcal{A}_4^{(L)}(1,4,2,3)
\end{eqnarray}
where $T^i$ denote the trace combinations of $SU(N_c)$ generators in the fundamental represen\-tation 
\beqa
T^1&=&Tr(T^{a_1}T^{a_2}T^{a_3}T^{a_4})+Tr(T^{a_1}T^{a_4}T^{a_3}T^{a_2}),\nonumber \\
T^2&=&Tr(T^{a_1}T^{a_2}T^{a_4}T^{a_3})+Tr(T^{a_1}T^{a_3}T^{a_4}T^{a_2}),\\
T^3&=&Tr(T^{a_1}T^{a_4}T^{a_2}T^{a_3})+Tr(T^{a_1}T^{a_3}T^{a_2}T^{a_4}).\nonumber
\eeqa

The four-point tree-level amplitude is always factorized, which is obvious within the superspace formalism. Hence, the colour decomposed L-loop amplitude can be represented 
 using the standard Mandelstam variables as
\beq
\mathcal{A}_4^{(L)}(1,2,3,4)=\mathcal{A}_4^{(0)}(1,2,3,4)M_4^{(L)}(s,t)
\eeq

\begin{figure}[h]
$M_4^{(L)}(s,t)$\vspace{-1cm}
\begin{center}
\includegraphics[scale=0.30]{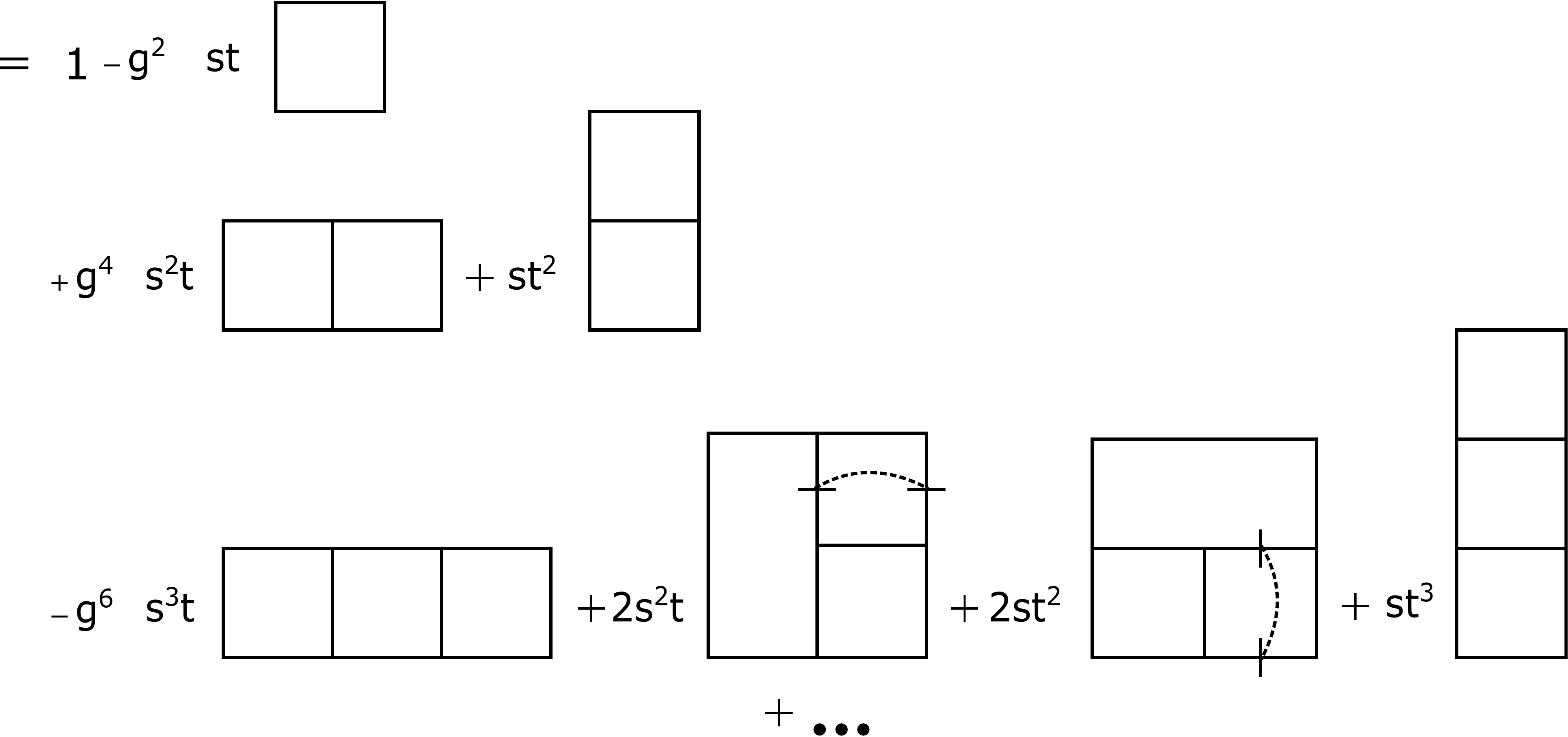}
\caption{The one-, two- and there-loop diagrams contributing to the four-point scattering amplitude in the SYM theory\label{expan}}
\end{center}\end{figure}

The factorized amplitude $M_4^{(L)}(s,t)$  can be expressed in terms of some combination of pure scalar master integrals times some polynomial in the Mandelstam variables shown in Fig.~\ref{expan}~\cite{Bern:2005iz}, which is universal for $D=4,6,8,10$ dimensions~\cite{spin-helicity}

To be more specific, we  further consider the case of $D=8$. Then,  explicit calculation of the diagrams gives the following result for the singular part of the amplitude in the two-loop order~\cite{K1}
 \beq
M_4^{(2)}(s,t)= 1-\frac{g_B^2 st}{3!\epsilon}(\mu^2)^\epsilon-\frac{g_B^4 st}{3!4!}\left(\frac{s^2+t^2}{\epsilon^2}+\frac{27/4 s^2+1/3 st +27/4 t^2}{\epsilon}\right)(\mu^2)^{2\epsilon} + ...  \label{gamma}
\eeq

Now we  apply eqs. (\ref{mult},\ref{coupling}) to get the  finite answer.
In the one-loop order, the coupling does not change   $g^2_{bare}=\mu^\epsilon g^2$
and the renormalization constant is chosen in the form 
\beq
Z_4^{(1)}=1+\frac{g^2 st}{3!\epsilon}
\eeq
that cancels the one-loop UV divergence.  Notice that the renormalization constant is not really a constant but depends on the kinematic factors $s$ and $t$.

In fact, this means that  this way we build an induced higher derivative theory where higher terms appear order by order of PT with fixed coefficients. For instance, the one loop term
$g^2 s t/\epsilon$ generates the gauge invariant counter term 
$$ \frac{g^2}{\epsilon}D_\rho D_\lambda F_{\mu\nu}D_\rho D_\lambda F_{\mu\nu},$$
that contains higher derivatives as well as new vertices with extra gauge fields, etc. 

In the two-loop order, the coupling changes now according to (\ref{coupling}), namely,
\beq
g^2_{bare}=\mu^\epsilon g^2(1+\frac{g^2 st}{3!\epsilon}) \label{1lcoup}\eeq
\\
 and the renormalization constant is taken in the form
 \beq
Z_4^{(2)}=1+\frac{g^2 st}{3!\epsilon}+\frac{g^4 st}{3!4!}\left(\frac{A_2s^2+B_2st+A_2t^2}{\epsilon^2}+
\frac{A_1s^2+B_1st+A_1t^2}{\epsilon}\right),\label{2lz}
\eeq
where the coefficients $A_i$ and $B_i$  have to be chosen in a way to cancel all divergences, both local and nonlocal ones. 

\begin{figure}[!ht]
\begin{center}
\includegraphics[scale=0.5]{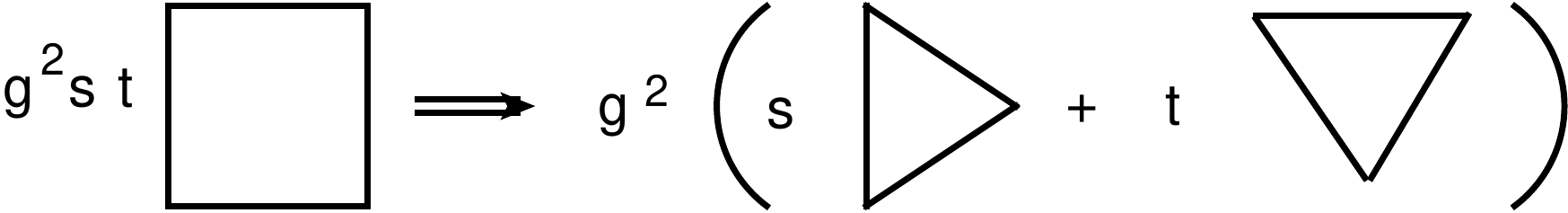}
\caption{Action of the Z-operator at the two-loop level}
\label{action}
\end{center}
\end{figure}

Multiplying the bare amplitude (\ref{gamma}) by the the renormalization constant (\ref{2lz})  and replacing  the bare coupling according to eq.(\ref{1lcoup}), one can notice that the replacement of $g^2_{bare}$  in the one loop term ($\sim g^2$) and multiplication of one-loop contributions from the renormalization constant $Z_4$  and from the amplitude ${\A}_4$ have the effect of subtraction of subdivergences in the two-loop graph. This is exactly what guarantees the locality of the counter terms within the $\R$-operation. However, contrary to the renormalizable case, here the renormalization constant contains the kinematic factors, the powers of momenta, which are external momenta for the subgraph but are internal ones for the whole diagram. Evaluating the counter-term, these kinematic factors have to be inserted inside the remaining diagram and integrated out. To be specific, we consider the corresponding term which appears when multiplying the one loop Z factor by the one loop amplitude. The $s$ and $t$ factors from the Z factor have to be inserted into the box diagram,
as shown in Fig.\ref{action}.

\begin{figure}[!ht]
\begin{center}
\includegraphics[scale=0.42]{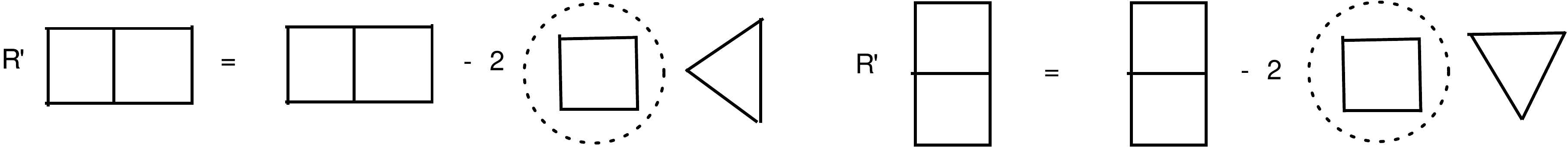}
\caption{$\R'$-operation for the two-loop diagrams}\label{2loop}
\end{center}
\end{figure}

This means that the usual {\it multiplication} procedure is modified: the Z factor becomes the {\it operator} acting on the diagram that inserts the powers of momenta into the diagram~\cite{K1}. This looks a bit artificial but exactly reproduces the $\R$-operation for the two-loop diagram shown 
in Fig.\ref{2loop}.

Inserting eqs.(\ref{2lz},\ref{1lcoup}) into eq.(\ref{mult}) and having in mind that 
\beq
sTriangle=-\frac{s}{4!\epsilon}(1+\frac{19}{6}\epsilon), \ \  tTriangle=-\frac{t}{4!\epsilon}(1+\frac{19}{6}\epsilon),
\eeq
one gets
\beqa
\bar{\A}_4& = &Z_4(g^2) \bar{\A}_4^{bare}|_{g^2_{bare}\mapsto g^2Z}\nonumber \\
&=& 1-\frac{g^2 \mu^\epsilon st}{3!\epsilon}+\frac{g^2 st}{3!\epsilon}-\frac{g^4 \mu^{2\epsilon} st}{3!4!}\left(\frac{s^2+t^2}{\epsilon^2}+\frac{27/4 s^2+1/3 st +27/4 t^2}{\epsilon}\right)  \\
&+& 2\frac{g^4 st}{3!\epsilon} \mu^{\epsilon}\frac{s^2+t^2}{4!\epsilon}(1+\frac{19}{6}\epsilon)+
\frac{g^4 st}{3!4!}\left(\frac{A_2s^2+B_2st+A_2t^2}{\epsilon^2}+
\frac{A_1s^2+B_1st+A_1t^2}{\epsilon}\right).\nonumber
\eeqa

One can see that the one-loop divergences ($\sim g^2$) cancel and the cancellation of the two-loop ones requires
\beqa
\frac{1}{\epsilon^2}: && -\frac{s^2+t^2}{3!4!}st +2 \frac{s^2+t^2}{3!4!}st + \frac{A_2s^2+B_2 st +A_2 t^2}{3!4!}st=0,\nonumber \\
\frac{\log\mu}{\epsilon}: &&  -2\frac{s^2+t^2}{3!4!}st +2 \frac{s^2+t^2}{3!4!}st =0,\nonumber \\
\frac{1}{\epsilon}: &&-\frac{st}{3!4!}(\frac{27}{4} s^2+\frac 13 st +\frac{27}{4} t^2)+2\frac{st}{3!4!}(s^2+t^2)\frac{19}{6}
+\frac{st}{3!4!}(A_1s^2+B_1st+A_1t^2)=0. \nonumber
\eeqa
One deduces that $A_2=-1,B_2=0, A_1=\frac{5}{12}, B_1=\frac 13$, so that the renormalization constant $Z_4$ takes the form~\cite{K1}
\beq
Z_4=1+\frac{g^2 st}{3!\epsilon}+\frac{g^4 st}{3!4!}\left(-\frac{s^2+t^2}{\epsilon^2}+
\frac{5/12 s^2+1/3st+5/12 t^2}{\epsilon}\right),\label{2lzp}
\eeq
which exactly corresponds to the one obtained using eq.(\ref{ZZ}). This expression now has to be substituted into eq.(\ref{coupling}) to obtain the renormalized  coupling. Note that it also depends on kinematics. 

\begin{figure}[ht]
\begin{center}
\includegraphics[scale=0.5]{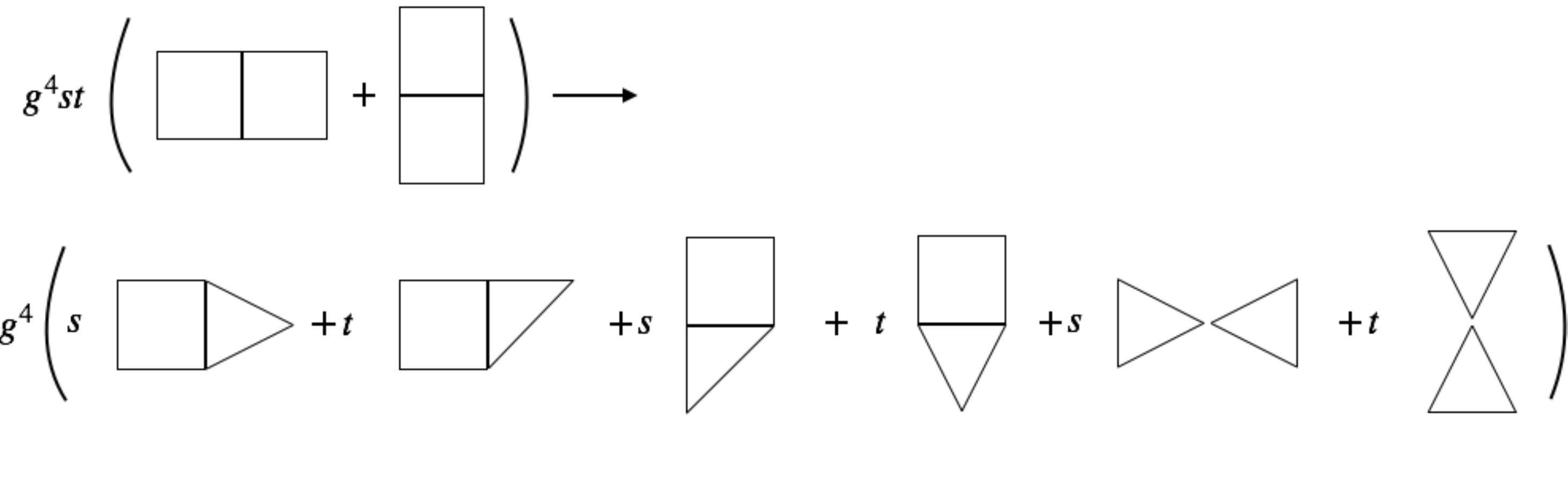}
\caption{Action of the Z-operator at the three-loop level. The first, second and the last two diagrams in the r.h.s correspond to the three loop box counter-terms and the third and fourth ones to the tennis-court counter-terms.}\label{3loop}
\end{center}
\end{figure}

The same way one can trace the action of the Z-operator in the three-loop diagram, as is shown in Fig.\ref{3loop}. In this case, besides the  3-loop box diagram one also has the tennis-court one, and the resulting counter-terms correspond to both of them.

\section{Normalization and the problem of arbitrariness}

We now turn to the problem of arbitrariness in subtraction of UV divergences which is the stumbling block in non-renormalizable interactions. We consider this problem in the context of transition from the minimal to non-minimal subtraction scheme. 

Remind first what happens in renormalizable theories. At the-one loop order, after removing the UV divergence, one is left with arbitrary subtraction constant $c_1$. Transition from the minimal to non-minimal scheme is achieved by multiplication of the amplitude by the finite renormalization constant
\beq
z=1+g c_1
\eeq
and the corresponding finite change of the coupling $g'=zg$.  To fix the value of $c_1$, one fixes the normalization of the vertex operator ($\phi^4$ in the case of the $g\phi^4$ theory, $\partial A  AA$ or $AAAA$ in the case of a gauge theory).  

This procedure repeats itself in each loop resulting in an infinite series of arbitrary subtraction constants $c_i$ which, however, are absorbed into a single renormalization constant
\beq
z=1+g c_1+g^2c_2+..., \ \ \ \  g'=g z.  \label{scheme}
\eeq
This constant normalizes a single coupling $g$ linked to a single operator.\footnote{If there are several operators with the same coupling, like the triple and quartic terms in gauge theories, they are related to each other and there is still one independent coupling. If two operators are truly independent, like $\phi^4$ and $\bar \psi \phi \psi$ in the Yukawa theory, then one has two independent renormalization constants and two couplings.}
 Fixing the normalization of this operator, we fix arbitrariness in the subtraction procedure by imposing a single condition.

Now, let us see what happens in the non-renormalizable case. Once again, one has an infinite set of  arbitrary constants at each order of perturbation theory.  And again, formally, they can be absorbed into a single renormalization constant. However, this constant happens to be  field and momentum dependent. This means that acting at the original operator it generates the whole infinite series of new operators order by order of PT. If one chooses to fix the normalization of each operator, which is the usual prescription of the renormalization procedure, one gets infinite arbitrariness with an infinite number of  imposed constraints. This is where we suggest deviation from the usual procedure and do not require normalization of each Green function associated with  these operators but require normalization of the amplitude as a whole \cite{K4}. This is achieved by imposing of a single condition just as in the renormalizable case. The amplitude will depend on all subtraction constants $c_i$ and will get contributions from all operators. At the same time, it will have a single coupling defined in a given subtraction scheme. When going from one scheme to another, one has to use the finite renormalization (\ref{scheme}) which does not change the normalization of one operator, as it happens in the renormalizable case, but a whole sequence of operators appearing in a given order of PT.
Therefore, it is not a simple change of a single coupling but of a whole infinite series of higher derivative terms. Thus, the whole arbitrariness is accumulated in one renormalization constant evaluated order by order in PT, which acts as an operator and generates an infinite series of terms.

Defining the amplitude, we fix the normalization of the coupling $g$ which,  
as was already mentioned, does not belong to a single operator but to an infinite series of operators. So fixing the amplitude, we simultaneously fix the whole series. Changing the normalization of any of these operators results in the corresponding change of the whole amplitude, which is the only relevant condition that we impose.  Thus, for example, changing the normalization of the $\phi^6$ operator in the $\phi^4$ theory, we change the four-point amplitude and this will correspond to another renormalization scheme. For a given choice of the subtraction constants of all the relevant operators one has the corresponding coupling $g$. In the renormalizable case this is called the scheme dependence.
In general, this dependence has a broader meaning.  Note that to fix arbitrariness, one can choose any amplitude.  Particular choice defines the coupling $g$ once and forever, and one has to use it in all
the other amplitudes. However, this choice is arbitrary and the transition to another one is nothing more than a scheme dependence. 

We discussed some features of the scheme dependence  in \cite{Sym}. It is governed by the finite renormalization (\ref{scheme}) where $z$ acts as an operator in the above mentioned sense.

\section{Recurrence relations and RG equations}

To show that the advocated construction is not just a declaration, we show below how one can effectively calculate the higher order asymptotics of the scattering amplitudes in non-renormalizable theories in much the same way as in renormalizable ones. We construct recurrence relations that connect the leading, subleading, etc.,
UV divergences (higher energy asymptotics) in subsequent orders of PT and then obtain a generalization of the renormalization group equations. These equations reflect the group structure of the proposed renormalization procedure when the renormalization constants are field and momentum dependent.

Any local  QFT has the property that in higher orders of PT after subtraction of divergent subgraphs,
i.e.   after performing the incomplete ${\cal R}$-operation (${\cal R}^\prime$-operation), the remaining UV divergences are local functions in the coordinate space or at maximum are polynomials of external momenta in momentum space. This follows from a rigorous proof of the Bogolyubov-Parasiuk-Hepp-Zimmermann $\R$-operation~\cite{BP,BPHZ} and is equally valid in non-renormalizable theories as well. 

This property allows one to construct recurrence relations that connect the divergent contributions in all orders of perturbation theory (PT) with the lower order ones.  In renormalizable theories these relations are known as pole equations (within dimensional regularization) and are governed by the renormalization group \cite{tHooft}. The same is true, though technically is more complicated, in any local theory,  as we have demonstrated in \cite{we}-\cite{we3} (see also review in \cite{Sym}).  We remind here some features of this procedure.

 Applying the ${\cal R}^\prime$-operation to a given graph $G$ in the n-th order of PT, one gets a series of divergent contributions shown below :
\beqa
{\cal R'}G_n&=&\frac{\A_n^{(n)}(\mu^2)^{n\epsilon}}{\epsilon^n}+\frac{\A_{n-1}^{(n)}(\mu^2)^{(n-1)\epsilon}}{\epsilon^n}+ ... +\frac{\A_1^{(n)}(\mu^2)^{\epsilon}}{\epsilon^n}\nonumber \\
&+&\frac{\B_n^{(n)}(\mu^2)^{n\epsilon}}{\epsilon^{n-1}}+\frac{\B_{n-1}^{(n)}(\mu^2)^{(n-1)\epsilon}}{\epsilon^{n-1}}+ ... +\frac{\B_1^{(n)}(\mu^2)^{\epsilon}}{\epsilon^{n-1}} \nonumber \\
&+&\mbox{lower\ pole\ terms,}\label{Rn}
\eeqa
where the terms like $\frac{\A_{k}^{(n)}(\mu^2)^{k\epsilon}}{\epsilon^n}$  or $\frac{\B_{k}^{(n)}(\mu^2)^{k\epsilon}}{\epsilon^{n-1}}$ originate from the $k$-loop graph which remains after subtraction of the $(n-k)$-loop counter-term.
The resulting expression has to be local and hence does not contain terms like $\log^l{\mu^2}/\epsilon^k$ from any $l$ and $k$. This requirement leads to a sequence of $n-1$ relations for $\A_i^{(n)}$,  $n-2$  ones for  $\B_i^{(n)}$, etc.,  which can be solved in favour of the lowest order terms
\beqa
\A_n^{(n)}&=&(-1)^{n+1}\frac{\A_1^{(n)}}{n}, \label{rela}\\
\B_n^{(n)}&=&(-1)^n \left(\frac 2n \B_2^{(n)}+\frac{n-2}{n}\B_1^{(n)}\right) \label{relb}.
\label{abc_coeff}
\eeqa

It is also useful to write down the local expression for the ${\cal KR'}$ terms (counter-terms) equal to
\beq
{\cal KR'}G_n=\sum_{k=1}^n \left(\frac{\A_k^{(n)}}{\epsilon^n} +\frac{\B_k^{(n)}}{\epsilon^{n-1}}+\cdots \right)\equiv
\frac{\A_n^{(n)'}}{\epsilon^n}+\frac{\B_n^{(n)'}}{\epsilon^{n-1}}+ \cdots.
\eeq
Then, one has, respectively,\vspace{-0.3cm}
\beqa
\A_n^{(n)'}&=&(-1)^{n+1}\A_n^{(n)}=\frac{\A_1^{(n)}}{n}, \label{relap} \\
\B_n^{(n)'}&=& \left(\frac{2}{n(n-1)} \B_2^{(n)}+\frac{2}{n}\B_1^{(n)}\right) \label{relbp}.
\eeqa

This means that performing the  ${\cal R}'$-operation in order to extract the leading pole, one can only take care of the one-loop  diagrams that survived after contraction and get the desired leading pole terms via eq.(\ref{rela}).  They can be calculated  in all loops purely algebraically starting from the one loop term $\A_1^{(1)}$. The same is true for subleading poles  but one should take into account the diagrams with two loops via eq.(\ref{relb}) just as it takes place in renormalizable theories \cite{KV}.   Here we restrict ourselves  to the leading  poles only. 

\begin{figure}[h]\vspace{0.3cm}
\begin{center}
\includegraphics[scale=0.28]{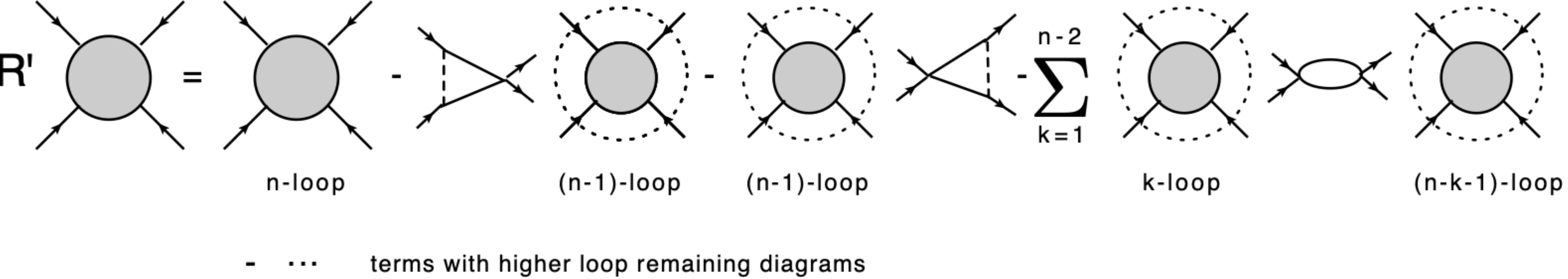}
\end{center}\vspace{-0.2cm}
\caption{The $\R'$-operation forу the 4-point function.  To simplify the picture, only the s-channel diagrams are shown}\label{Rprime}
\end{figure}

\begin{figure}[h]
\begin{center}
\includegraphics[scale=0.38]{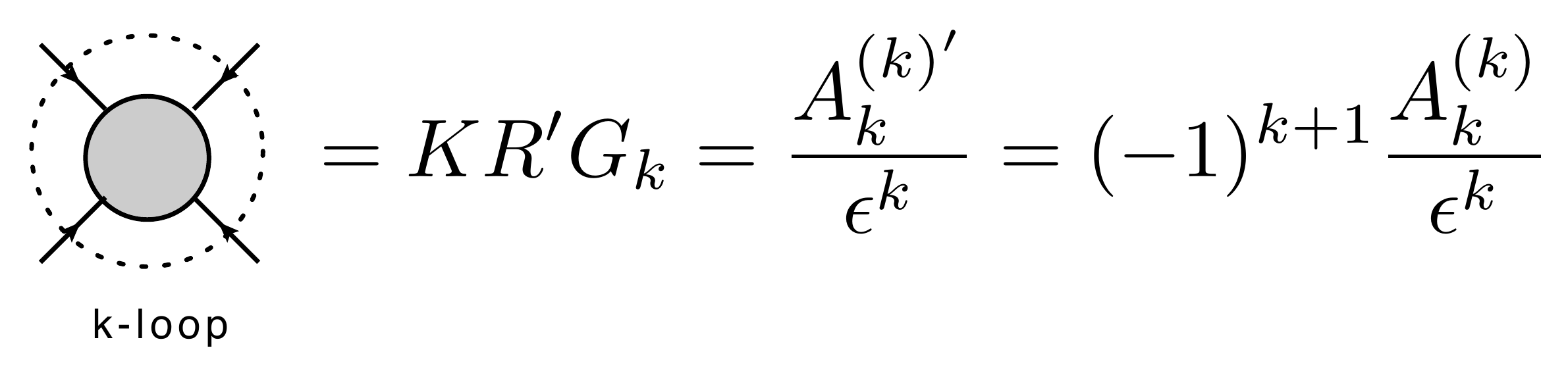}
\end{center}\vspace{-0.5cm}
\caption{The counter-term $K\R'\ G$. The leading divergence is shown} \label{kr'}
\end{figure}

In what follows, we consider the $2 \to 2$ scattering amplitude on shell and take the massless case. This means that all $p_i^2=0$ and the amplitude depends on the Mandelstam variables
$s,t,u$ with $s+t+u=0$. 
The $\R'$-operation for the 4-point function is shown schematically in Fig.\ref{Rprime}, where
the dotted line denotes the counter term obtained by the action of the $\R'$-operation on the corresponding subgraph (see Fig.\ref{kr'}). 


It should be stressed that the usual BPHZ procedure applied to the Green function includes the off-shell counter terms. However, if we require finiteness of the on-shell amplitudes, only the on-shell counter terms contribute. At first sight it seems that calculating the higher order diagrams, one will have divergent subgraphs with off-shell legs and therefore will require the off-shell counter terms.  But  this is not so. Since the counter terms are local, i.e. are polynomial of momenta, the off-shell parts will be proportional to $p_i^2$ where $p_i$ are momenta flowing through the external leg of the diagram
(the on-shell condition is $p_i^2=0$). These terms will cancel the corresponding internal line of the remaining diagram and as a result this  diagram disappears on-shell.

The  action of the $\R'$-operation shown in Fig.\ref{Rprime} is almost universal for any theory. The specific feature of particular interaction manifests itself in the first two terms which contain the live loop on the left or right edge of the diagram. In the case of triple vertices it is a triangle while for quartic vertices it is a bubble.   If both the vertices are present, one has both contributions that reflect the diagrams appearing in a given theory. 
The nonlinear term that contains the live loop in the middle is always a bubble. We also remind the reader that the only diagrams that give a contribution to the $n$-th order pole at $n$ loops are those that contain divergent subgraphs from 1 to $(n-1)$ loops.

Thus, if one is interested in the leading poles in the n-th loop, everything is reduced to the one-loop diagrams that survived after contraction of the n-1 loop subgraphs.  Then, for the four-point amplitude equation (\ref{rela}) leads to the following recurrence relation \cite{K3,K4}:

\beq
\includegraphics[scale=0.40]{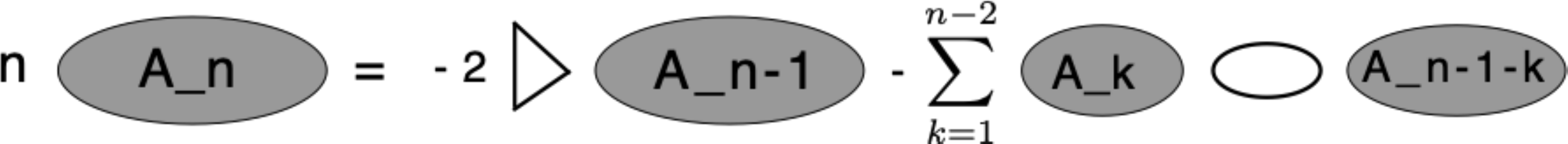},\label{recgraf}
\eeq
where the dark circles represent the singular parts $\A_k^{(k)'}$ but due to eq.(\ref{relap}) can be replaced by $\A_k^{(k)},\  k=1,..,n$. 

This recurrence relation  allows one to calculate all the leading divergences $\A_n^{(n)}$ starting from the one-loop term $\A_1^{(1)}$.   Below we demonstrate explicit realization of the recurrence relation
(\ref{recgraf}) using the theory $g \phi^4_D$ and the SYM$_D$ theory  for $D=4,6,8,10$ as an example.

To sum the leading divergences (or the leading logarithms in the scattering amplitude which is the same),
one has to solve this recurrence relation. This task in general seems to be impossible. Instead, one can 
convert the recurrence relation into a differential equation for the sum of all terms
\beq
\A(z)=\sum_{n=1}^\infty (-z)^n \A_n^{(n)}. \label{sum}
\eeq 
Multiplying eq.(\ref{recgraf}) by  $(-z)^n$  and taking the sum over $n$ from 2 to $\infty$, one gets the differential equation for the function  $\A(z)$ (\ref{sum}) which  can be symbolically written as
\beq
\frac{d}{dz} \A(z)=-1-2\int_{\rotatebox{90} {$\bigtriangledown$}}\A(z)-\int_{\rotatebox{90} O}\A^2(z),  \label{rgeq}
\eeq
where integration is performed over the remaining one-loop diagrams. 

Equation (\ref{rgeq})  is nothing else than the renormalization group equation. Below we present it in an explicit form for some particular cases. Note that in the renormalizable case, all the integrals disappear  and one has an ordinary differential equation while in the non-renormalizable case it is integro-differential.

Let us stress once more that in the renormalizable case the $\R$-operation is reduced to multiplication of the amplitude by the renormalization constant $Z$ and the corresponding multiplication of the coupling
(see eq.(\ref{coupling}), so that the group operation is simply a multipli\-cation. In the non-renormalizable case, the $\R$-operation is not reduced to multipli\-cation, but the group character 
of renormalization remains untouched. The only difference is that one just has no inverse operation anymore due to integration in eq.(\ref{rgeq}), so it looks like one has a semi-group rather than a group in this case.

\section{Illustration}

We demonstrate now how the  recurrence relations for the leading poles  (\ref{recgraf})  and the RG equations (\ref{rgeq}) are written explicitly in some particular cases.

\subsection{$g \phi^4_D$ Theory}
As the first example we consider the $g\phi^4$ theory in dimensions $D=4,6,8,10$~\cite{K2}.

Define the four-point function $\Gamma_4$  as follows:
\beq
\Gamma_4(s,t,u)= g  \bar \Gamma_4(s,t,u)=g(1+\Gamma_s(s,t,u)+\Gamma_t(s,t,u)+\Gamma_u(s,t,u)),
\eeq
where the functions $\Gamma_t(s,t,u)$ and $\Gamma_u(s,t,u)$ are related to $\Gamma_s(s,t,u)$
by the cyclic change of arguments;  $\Gamma_4$ obeys the PT loop expansion over $g$
\beq
\Gamma_s=\sum_{n=1}^\infty (-z)^n S_n, \ \ \Gamma_t=\sum_{n=1}^\infty (-z)^n T_n, \ \ \Gamma_u=\sum_{n=1}^\infty (-z)^n U_n, \ \ \ z\equiv \frac g\epsilon,
\eeq
where we keep only the leading pole terms.

To calculate them, we use the power of eq.(\ref{rela}). Indeed, in our notation $\A^{(n)}_n=S_n+T_n+U_n$ and can be expressed through $\A^{(n)}_1$ which is given by the diagrams shown in Fig.\ref{Rprime}. The dotted diagrams corresponding to $\K\R'G_k$ are given by $\A_k^{(k)'}$ and are also expressed through $\A^{(k)}_1$, according to eq.(\ref{relap}).

Then for the scattering amplitude 
  $S_n$  one has the  recurrence relation (\ref{recgraf}) where the integration over the one-loop subgraph can be carried out in general form introducing the Feynman parameters. The result is the following:
\beqa
&&n S_n(s,t,u)\nonumber\\
&=&\frac{s^{D/2-2}}{\Gamma(D/2-1)}\! \!\int_0^1 \!\! \! dx [x(1\!-\! x)]^{D/2-2}\left( S_{n-1}(s,t',u')\!+\! T_{n-1}(s,t',u')\!+\! U_{n-1}(s,t',u')\right)\nonumber\\
&+&\frac 12 \frac{s^{D/2-2}}{\Gamma(D/2-1)}\! \!\int_0^1 \!\! \! dx [x(1\!-\! x)]^{D/2-2}\sum_{k=1}^{n-2}\sum_{p=0}^{(D/2-2)k}\sum_{l=0}^{p}\frac{1}{p!(p+D/2-2)!}\times \label {rec}\\
&\times&\frac{d^p}{dt'^l du'^{p-l}}(S_k+T_k+U_k)\frac{d^p}{dt'^l du'^{p-l}}(S_{n-k-1}+T_{n-k-1}+U_{n-k-1})s^p[x(1-x)]^p t^l u^{p-l}\nonumber
\eeqa
where $t'=- xs, u'=-(1-x)s$. And the same for the other partial amplitudes  $T_n$ and  $U_n$ with the cyclic change of arguments.

The first linear term of eq.(\ref{rec}) corresponds to the first two diagrams in Fig.\ref{Rprime} and the  second nonlinear term is due to the third diagram  with the  live loop in the middle. Integration over $x$ is just integration over the Feynman parameter in the loop diagram. Multiple sums appear due to the $g_{\mu\nu}$ factors that arise when integrating multiple momenta in the numerator of the diagrams.
This recurrence relation allows one to calculate all the leading divergences in all loops in a pure algebraic way starting from the one-loop diagram.

One can convert the recurrence relation (\ref{rec}) into a differential equation for the function $\Gamma_s(s,t,u|z)$ taking the sum over $n$ of eq.(\ref{rec}). Thus, taking the sum  $\sum_{n=2}^\infty (-z)^{n-1} $, one gets
\beqa
&-&\frac{d \Gamma_s(s,t,u)}{dz}=\frac 12 \frac{\Gamma(D/2-1)}{\Gamma(D-2)}s^{D/2-2}\nonumber \\ &+&\frac{s^{D/2-2}}{\Gamma(D/2-1)}\! \!\int_0^1 \!\! \! dx [x(1\!-\! x)]^{D/2-2}\left[\Gamma_s(s,t',u')\!+\! \Gamma_t(s,t',u')\!+\! \Gamma_u(s,t',u')\right]\vert_{\scriptsize  \begin{array}{l}
t'=- xs, \\ u'=-(1-x)s\end{array}}\nonumber\\
&+&\frac 12 \frac{s^{D/2-2}}{\Gamma(D/2-1)}\! \!\int_0^1 \!\! \! dx [x(1\!-\! x)]^{D/2-2}\sum_{p=0}^{\infty}\sum_{l=0}^{p}\frac{1}{p!(p+D/2-2)!}\times \label {eqs}\\
&\times&\left(\frac{d^p}{dt'^l du'^{p-l}}(\Gamma_s+\Gamma_t+\Gamma_u)\vert_{\scriptsize  \begin{array}{l}
t'=- xs, \\ u'=-(1-x)s\end{array}}\right)^2 s^p[x(1-x)]^p t^l u^{p-l}, \nonumber
\eeqa
with the boundary condition $\Gamma_s(z=0)=0$
and the same for $\Gamma_t$ and $\Gamma_u$ with the cyclic change of arguments. 

Equation (\ref{eqs}) can be simplified if written for the whole function $\bar \Gamma_4$.  One can also notice that due to the one-to-one correspondence between $1/\epsilon$ and $\log\mu^2$  one can rewrite the equation for $\bar \Gamma_4$ in a more familiar way
\beqa
&&\frac{d \Gamma_s(s,t,u)}{d\log\mu^2}=
-\frac{g}{2} \frac{s^{D/2-2}}{\Gamma(D/2-1)}\! \!\int_0^1 \!\! \! dx [x(1\!-\! x)]^{D/2-2}\sum_{p=0}^{\infty}\sum_{l=0}^{p}\frac{1}{p!(p+D/2-2)!}\times \nonumber\\
&&\times \left(\frac{d^p \bar \Gamma_4(s,t',u')}{dt'^l du'^{p-l}}\vert_{\scriptsize  \begin{array}{l}
t'=- xs, \\ u'=-(1-x)s\end{array}}\right)^2 s^p[x(1-x)]^p t^l u^{p-l},
 \label {eqrg}
\eeqa
with the boundary condition $\Gamma_s(\log\mu^2=0)=0$. 

Equation (\ref{eqrg}) is nothing else than the desired generalized RG equation for the $\phi^4_D$ theory in $D$-dimensions. To see it, consider the case when $D=4$. It corresponds to a well-known renormalizable theory where the divergent part of the amplitude $\bar \Gamma_4$ does not depend on $s,t$ and $u$ and hence one can drop the integrals and sums  in eq.(\ref{eqrg}). Adding the terms with $\Gamma_s, \Gamma_t$ and $\Gamma_u$ together, one has
\beq
D=4: \ \ \ \ \ \ \  \frac{d\bar \Gamma_4}{d\log\mu^2}=
-\frac 32 g \bar \Gamma_4^2, \ \ \ \ \bar \Gamma_4(\log\mu^2=0)=1. \label{d4}
\eeq

To find the high energy behaviour of the amplitude $\Gamma_4$ in the fixed angle regime when $s\sim t\sim u\sim E^2$, one has to solve eq.(\ref{eqrg}).   In the case of $D=4$, eq.(\ref{d4}) has
 an obvious solution in the form of a geometrical progression
\beq
\bar \Gamma_4=\frac{1}{1+\frac 32g \log(\mu^2/E^2)}\ \ \ \ \ \mbox{or}\ \ \ \ \    \Gamma_4=\frac{g}{1+\frac 32 g \log(\mu^2/E^2)}.
\eeq

This solution suggests the form of the solution to eq.(\ref{eqrg}) for arbitrary $D$.  It can be written as
\beq
\Gamma_4(s,t,u)={\mathcal P} \frac{g}{1+\frac 12 \frac{\Gamma(D/2-1)}{\Gamma(D-2)}g (s^{D/2-2}+t^{D/2-2}+u^{D/2-2})\log(\mu^2/E^2)},
\label{sol}
\eeq
where the symbol ${\mathcal P}$ means the ordering in a sense of eq.(\ref{rec}), i.e.  when expanding the geometrical progression in a series over $g$, one has to choose a single loop in the $s,t$ or $u$ channel and then integrate the powers of $s,t$ and $u$ over this loop. This gives exactly the PT series of the form (\ref{rec}).  Symbolically, one can write eq.(\ref{sol}) as
 \beq
\Gamma_4(s,t,u)={\mathcal P} \frac{g}{1+g \A_1^{(1)}\log(\mu^2/E^2)}
\label{sol2}.
\eeq
Perturbative expansion then looks like
\beq
{\mathcal P}g \sum_{n=0}^{\infty} (-g)^n \log^n(\mu^2/E^2) (\A_1^{(1)})^n,
\eeq
where the n-th term  has to be understood as
\beq
{\mathcal P}(\A_1^{(1)})^n = \int_0^1 dx \sum_{k=0}^{n-1}\  \overrightarrow{ {\mathcal P}(\A_1^{(1)})^k }\ \A_1^{(1)} \ \overleftarrow{{\mathcal P}(\A_1^{(1)})^{n-1-k}}, 
\eeq
where the arrow means that one has to integrate the expression under the arrow sign through 
$\A_1^{(1)} $ in a sense of eq.(\ref{rec}).

Solution (\ref{sol}) reproduces the PT series for the leading logarithms and allows one to find the high energy behaviour of the amplitude $\Gamma_4$, studying the singularities under the ${\mathcal P}$-ordering. Indeed, one can easily see that the sign of the logarithm in the denominator of eq.(\ref{sol}) 
is always positive (exception is the case of D=6 where due to the condition $s+t+u=0$  all the leading divergences cancel on shell). Thus, in the regime $E\to \infty$ one always has a Landau pole for any $D$ just as it takes place for $D=4$.

\subsection{SYM$_D$ Theory}

Consider now the supersymmetric Yang-Mills theory in $D$ dimensions~\cite{we,we2,we3}. The difference from the $\phi^4$
theory is that one has triple vertices in this case and after shrinking the divergent subgraphs, one is left with the triangle one-loop diagrams rather than the bubble ones. Hence, the corresponding  Feynman parameters  are $x$ and $y$ in this case. 

Denote by $S_n(s,t)$ and $T_n(s,t)$ the sum of all contributions in the $n$th order of PT in the $s$ and $t$ channels, respectively, so~that
\beq
M_4(s,t)\bigg|_{\mbox{leading UV div.}}=\sum_{n=0}^{\infty}(-g^{2})^n\frac{S_n(s,t)+T_n(s,t)}{\epsilon^n},
\eeq
we get the following recurrence relations:
\beqa
&&nS_n(s,t)\nonumber\\
&=&-2 \frac{s^{D/2-2}}{\Gamma(D/2-2)} \int_0^1 dx \int_0^x dy\  [y(1\!-\!x)]^{D/2-3} \ (S_{n-1}(s,t')+T_{n-1}(s,t'))|_{t'=t(x-y)-sy}\nonumber \\ &+&
\frac{s^{D/2-2}}{\Gamma(D/2-3)} \int_0^1\!\! dx \ [x(1\!-\! x)]^{D/2-2} \sum_{k=1}^{n-2}  \sum_{p=0}^{(D/2-2)k-2} \frac{1}{p!(p+D/2-2)!} \times \label{recYM}\\
&&\times  \frac{d^p}{dt'^p}(S_{k}(s,t')+T_{k}(s,t'))  \frac{d^p}{dt'^p}(S_{n-1-k}(s,t')+T_{n-1-k}(s,t'))|_{t'=-sx} \ (tsx(1-x))^p,  \nonumber
\eeqa
and the same for  $T_n$ with the change of variable $s,t\ \leftrightarrow t,s$. As a starting value one has the one loop box  diagram equal to
\beq
S_1+T_1=Sing\  \Box_D=\frac{1}{\Gamma (D-4)}\sum_{k=0}^{D/2-4}k!(D/2-4-k)!s^kt^{D/2-4-k}.
\eeq
The leading divergences in any order of PT can be evaluated  in an algebraic form using these recurrence relations, starting from the known values of $S_1$ and $T_1$.

The case of D=6 is somewhat special since the box diagram is convergent here and the first UV divergence comes at three loops. This is the so-called tennis-court diagram shown in Fig.\ref{expan}.
As a resul,t the non-linear term in eq.(\ref{recYM}) disappears  and one has
\beq
D=6 \ \ \ \ nS_n=-2s\int_0^1 dx \int_0^x dy\   (S_{n-1}(s,t')+T_{n-1}(s,t'))|_{t'=t(x-y)-sy}
\eeq
with  $S_3=-\frac s3, \ T_3=-\frac t3$.

Similarly to the $\phi^4$ case, these recurrence relations include all the diagrams of a given order of PT and allow one to  sum all orders of PT. This can be done by multiplying both sides of equations (\ref{recYM}) by $(-z)^{n-1}$, where $z=\frac{g^2}{\epsilon}$ and summing up from $n = 2$ to infinity. Denoting~the sum by $\Sigma(s,t,z)=\sum_{n=1}^\infty S_n(s,t) (-z)^n$, we~finally obtain the following differential equation for the s-channel amplitude:
\beqa
&&\frac{d}{dz}\Sigma(s,t,z)=-S_1\\
&&+2 \frac{s^{D/2-2}}{\Gamma(D/2-2)} \int_0^1 dx \int_0^x dy\   [y(1\!-\!x)]^{D/2-3}\ (\Sigma(s,t',z)+\Sigma(t',s,z))|_{t'=tx+yu}\nonumber \\
&&-\frac{s^{D/2-2}}{\Gamma(D/2-3)}   \int_0^1\! dx \ [x(1\!-\! x)]^{D/2-2}  \sum_{p=0}^\infty \frac{1}{p!(p+D/2-2)!} \nonumber \\
&&(\frac{d^p}{dt'^p}(\Sigma(s,t',z)+\Sigma(t',s,z))|_{t'=-sx})^2 \ (tsx(1-x))^p. \nonumber
\eeqa
The same equation with the replacement $s \leftrightarrow t$ is valid for $\Sigma(t,s,z)=\sum_{n=1}^\infty T_n(s,t) (-z)^n$.

Again, the case of $D=6$ is special and one has the linear equation
 \beq
 \frac{d}{dz}\Sigma(s,t,z)=s-\frac{2}{z}\Sigma(s,t,z)+2s \int_0^1 dx \int_0^x dy\ (\Sigma(s,t',z)+\Sigma(t',s,z))|_{t'=xt+yu}.
\label{eqr}
\eeq

Contrary to the scalar theory considered above, there is no $D=4$ instructive case here since the $D=4$ $N=4$ supersymmetric theory is totally UV finite. We therefore have no simple solution to the RG equations like (\ref{sol}). Still, it is possible to get an analytic solution in some particular cases. For instance, for a sequence of ladder diagrams for D = 6 in the s -channel one has
\beq
\Sigma_L(s,z)=\frac{2}{s^2z^2}(e^{sz}-1-sz-\frac{s^2z^2}{2}).\label{lad6}
\eeq
 As follows from (\ref{lad6}), the amplitude exponentially increases in one direction and drops in the other. A similar behaviour follows from numerical analysis of the complete equation. In the (s,t) and (s,u) channels the amplitude is decreasing with energy, while in the (t,u) channel it is increasing exponentially. 
At D = 8, the sequence of ladder diagrams in the s -channel is described by the Riccati equation and has the form
\beq
\Sigma_L(s,z)=-\sqrt{5/3} \frac{4 \tan(zs^2/(8 \sqrt{15}))}{1 - \tan(zs^2/(8 \sqrt{15}))\sqrt{5/3}}.\label{lad8}
\eeq
The resulting function has an infinite periodic sequence of Landau poles. The numeric solution of the complete equation shows that this behaviour is characteristic of the general solution as well.
In the cases $D=8,10$,
the amplitude possesses infinite number of Landau poles for all channels \cite{we2,we3}.

One must stress that while the recurrence relations and the generalized RG equations can be written for any theory, in the non-renormalizable case they look quite complicated, though the symbolic form (\ref{rgeq})
is always valid. Solutions to the RG equations strongly depend on a theory but eventually are still defined by the one-loop term.

\section{Conclusion}

 Our main statement is that non-renormalizable theories are self-consistent, they can be well treated within the usual BPHZ  $\R$-operation.  In them, you can perform the procedure for eliminating divergences, fixing the existing arbitrariness, and suming up the leading asymptotics, just as it happens in renormalized theories. They differ from renormalizable ones only in their technical implementation.

 \section*{Acknowlegements}
This paper is supported by the Russian Scientific Foundation under grant  \# 16-12-10306.  The author is grateful to his collaborators  and members of the Bogoliubov Laboratory for numerous fruitful discussions.

\end{document}